\documentclass[useAMS,usenatbib,usegraphicx,letterpaper]{mn2e}
\usepackage{amssymb}
\usepackage{amsfonts}
\usepackage{amsbsy}
\usepackage{color}

\newcommand{\be}{\begin{equation}}
\newcommand{\ee}{\end{equation}}
\newcommand{\bea}{\begin{eqnarray}}
\newcommand{\eea}{\end{eqnarray}}

\newcommand{\ben}{\begin{eqnarray}}
\newcommand{\een}{\end{eqnarray}}

\title[Oscillations in the dark energy EoS: new MCMC lessons]{Oscillations in the dark energy equation of state: new MCMC lessons}
\author[R. Lazkoz,  V. Salzano, I. Sendra]{Ruth Lazkoz, Vincenzo Salzano, Irene Sendra\\
Fisika Teorikoaren eta Zientziaren Historia Saila, Zientzia eta Teknologia Fakultatea, \\ Euskal
Herriko Unibertsitatea, 644 Posta Kutxatila, 48080 Bilbao, Spain}

\begin{document}



\maketitle

\begin{abstract}

We study the possibility of detecting oscillating patterns in the
equation of state (EoS) of the dark energy using different
cosmological datasets. We follow a phenomenological approach
and study three different oscillating models for the EoS, one of them periodic
and the other two damped (proposed here for the first time).
All the models are characterised by the amplitude value, the
centre and the frequency of oscillations. In contrast to
previous works in the literature, we do not fix the value of the frequency
to a fiducial value related to the time extension of chosen
datasets, but consider a discrete set of values, so to avoid 
arbitrariness and try and detect  any possible time
period in the EoS. We test the models using
a recent collection of SNeIa, direct Hubble data and 
Gamma Ray Bursts data. Main results are: I. even if constraints on the
amplitude are not too strong, we detect a trend of it versus the
frequency, i.e. decreasing (and even negatives) amplitudes for
higher frequencies; II. the centre of oscillation (which
corresponds to the present value of the EoS parameter) is very well constrained,
phantom behaviour is excluded at $1\sigma$ level and 
trend which is in agreement with the one for the amplitude appears; III. the
frequency is hard to constrain, showing similar statistical
validity for all the values of the discrete set chosen, but the
best fit of all the scenarios considered is associated with a
period which is in the redshift range depicted by our cosmological
data. The ``best" oscillating models are compared with 
$\Lambda$CDM using dimensionally consistent a Bayesian approach based information criterion
and the conclusion reached is the non existence of significant evidence against dark energy oscillations.
\end{abstract}


\section{Introduction}

We have nowadays  a great amount of independent data sets available
for studying the present dynamical state of the Universe. High quality
data coming from the Hubble
diagram of Type Ia Supernovae (\cite{Riess04,ast05,clo05});
the measurements of cluster properties as the
mass, the correlation function and the evolution with redshift of
their abundance (\cite{eke98,vnl02,bach03,bb03});  the
optical surveys of large scale structure
(\cite{pope04,cole05,eis05}); the anisotropies of the cosmic
microwave background (\cite{Boom,WMAP}); the cosmic shear measured
from weak lensing surveys (\cite{vW01,refr03}) and the
Lyman\,-\,$\alpha$ forest absorption (\cite{chd99,mcd04}) are
evidences toward an \textit{apparently clear} picture of our
universe at present. It is characterised by: I. spatial flatness,
II. a subcritical matter content, III. and accelerated expansion.

But the \textit{clearness} of this sketch poses a more interesting
and deeper problem: how can we interpret all these features in the framework of
a self-consistent theoretical cosmological model? This is the main
task of modern cosmology and no unique answers have been given so far.

The $\Lambda$CDM model is the simplest (from a statistical point of view) and the most
accepted (it is called \textit{concordance model} just for this
reason), and in this scenario acceleration is
driven by the famous cosmological constant, $\Lambda$, which
contributes to the energy/matter content of the Universe by more
than a $74\%$. At the same time it requires the presence of a large
amount of \textit{cold dark matter} (about the $22\%$ of total
 energy/matter content), i.e. non-baryonic matter which does not interact
with electromagnetic radiation, but it is detectable only by its
gravitational interaction with ordinary baryonic matter (which makes the remaining
$4\%$). This
model provides a good fit to most of the data
(\cite{Teg03,Sel04,sanch05}) giving a reliable snapshot of the
current Universe; but it is also affected by serious
theoretical shortcomings that have motivated the search for
alternative candidates generically referred to as \textit{dark
energy} or \textit{quintessence}. Such models  range from conventional scalar
fields rolling down self-interaction potentials, to non-canonical scalar field models (phantom, k-essence,etc.);
from phenomenological unified models of dark energy and dark
matter to alternative gravity theories
(\cite{PB03,Pad03,Copeland}).

Unfortunately, so many data are not yet able to give us a
definitive answer about the origin and nature of the
acceleration of the Universe; not able to solve the \textit{coincidence problem}, (namely,
why we are currently detecting an energy-density for dark energy
which is quite equal to the one of dust matter); and not able to state
what the right EoS of dark energy is.

Regarding the EoS,  using current data one will be typically only able to infer that
the dark energy effective EoS parameter $w$\footnote{This is the
usual  factor which relates  pressure and
density of any given component through the relation $w_{X} \doteq p_{X}/\rho_{X}$.} is
close to $-1$. But any small deviation from this value could give
a different theoretical scenario: if it is \textit{exactly equal}
(when of course referring to observational errors) to $-1$, we
have a cosmological constant; if it is larger than $-1$, we have a
quintessence model; while if it is smaller than $-1$ we have the so
called phantom dark energy. In addition, the data seem to indicate that the fractional energy
densities of two main components of the Universe, i.e. dark matter and dark energy, are
very similar at present, and the label ``coincidence problem'' has been coined to refer to
this striking similarity.

One of the most interesting solutions proposed to try and throw some light on these
questions is \textit{oscillating dark energy}
(\cite{Dodelson00,Feng06,Sahni00}). Such a model can easily solve
the coincidence problem in a very natural way due to periods of
acceleration, and can be also used as a good candidate for the
unification of the late time acceleration (the one observed at present) with inflation (an early time
acceleration period). In this context we have to
underline the difference between assessing a periodic or
non-monotonic potential and an oscillating dark energy EoS. In many
cases such potentials do not give rise to a periodic $w$; one
example can be found in \cite{Frieman95}, where the proposed field is
clearly periodic but the derived $w$ can be well described by
the CPL parametrization for dark energy introduced in
\cite{Chevallier01} and \cite{Linder03}.

In this paper we are going to follow the method from
\cite{Linder06}, by examining some directly proposed phenomenological
periodic equations of state for the dark energy using different cosmological observations. Specifically, we
are going to set constraints on the location of the centre of the range about which $w$ oscillates and the amplitude
of the oscillations, and we will also constrain the fractional energy density of matter.

Actually, in the models to be considered
there is another important parameter, the frequency of the oscillation. Relevant though it is, leaving this parameter
completely free leads  to a high dimensionality statistical problem, but  given the precision of the data available at present
it seems that problems of that sort cannot be tackled satisfactorily; i.e. there seems to be not enough quality in the data to constrain more than two dark energy parameters \cite{Linder05,Sullivan08}. In the literature on oscillating
dark energy, the usual practice has been  choosing a specific single value of this oscillation frequency and stick with it. In order
to fix it one may resort to an argument by \cite{Linder06} which suggests the lowest bound on this frequency for the data
in use to be able to discriminate an oscillating behaviour from a monotonic one. We wish to carry out here a more thorough
study of oscillating dark energy than previous works, and to this end we choose a discrete set of values of the frequency (above and
below that bound) and then obtain constraints on the rest of parameters.  Relaxing assumptions on the frequency, as compared
to previous works, will allow us able to draw stronger conclusions; yet this is not the only novelty of our analysis, as will
be shown in what follows.

In this exploration of possible oscillating patterns in the expansion of the Universe induced
by the dark energy component we also find it interesting to consider non-periodic oscillating models. Specifically we deal with
scenarios which display a similar start off to the
popular oscillating model by Linder, but then depart from it as their amplitude gets smaller as $z$ grows and the distance between nodes tends to converge to a specific value. In this direction, we propose a pair of models in which dark energy oscillations are modelled via special functions. Comparison between models of that sort and the usual trigonometric parametrization provides hints about which features in the oscillations are favoured or disfavoured.

\begin{figure*}
\begin{minipage}{0.49\textwidth}
\includegraphics[width=.70\textwidth]{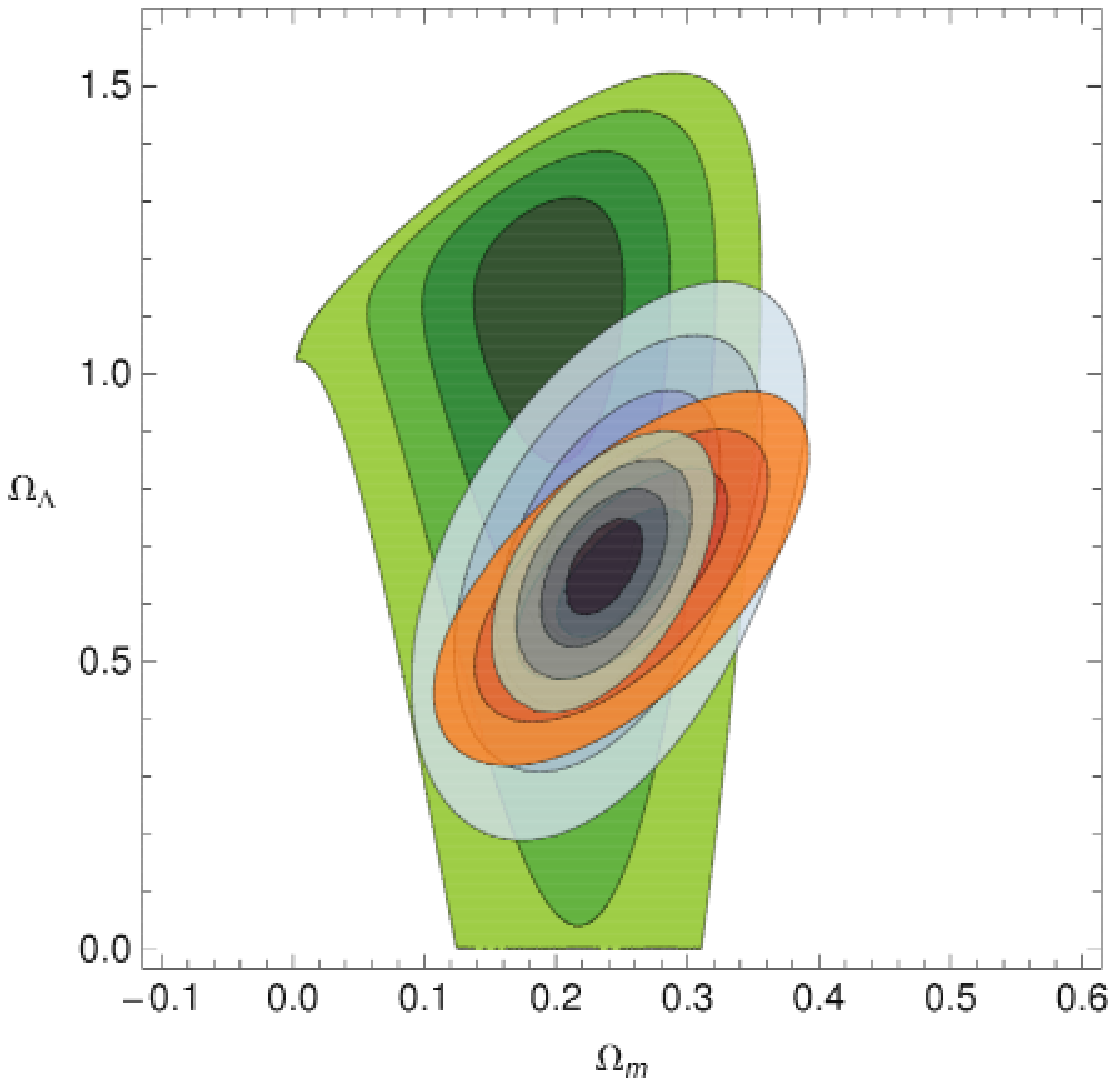}\\
\centering
(a) $\Lambda$CDM
\end{minipage}
\begin{minipage}{0.49\textwidth}
\includegraphics[width=.70\textwidth]{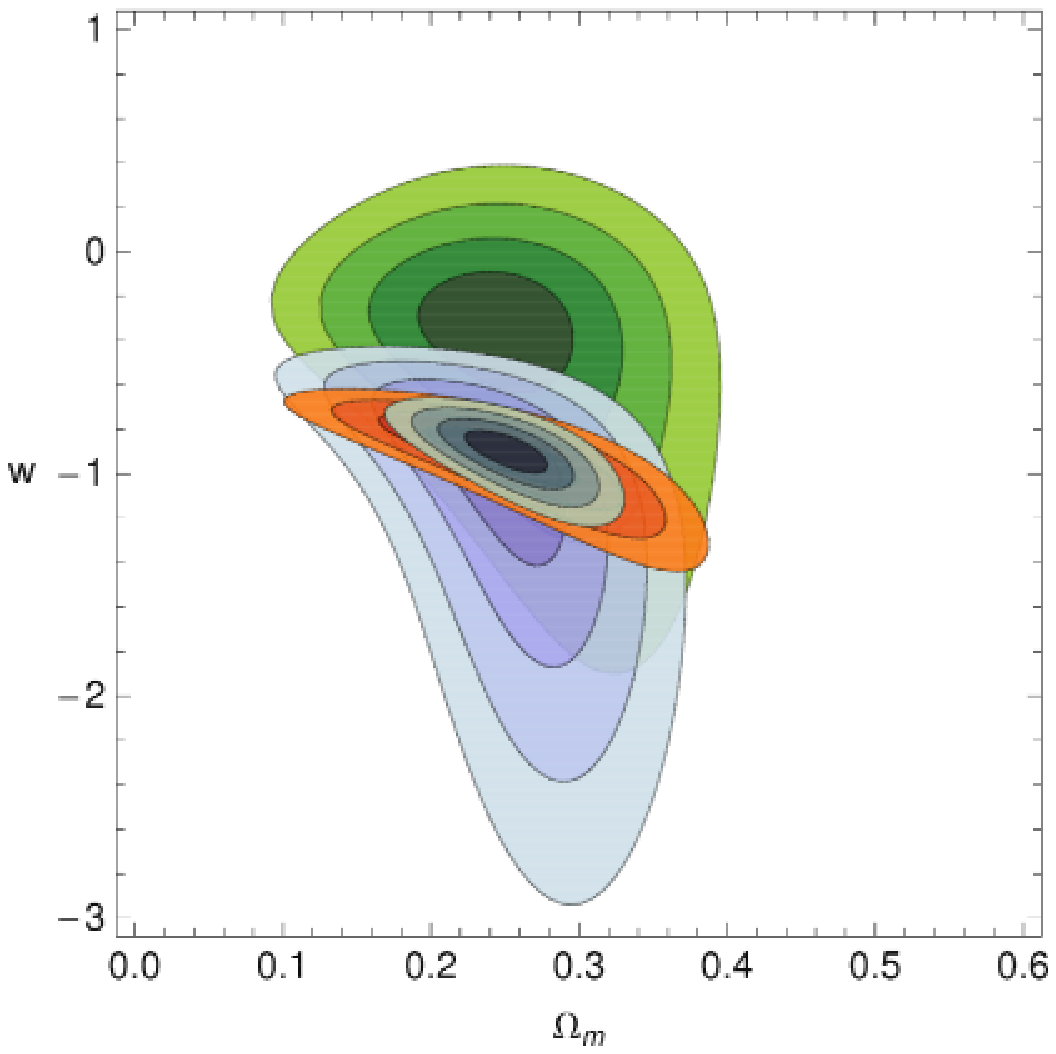}\\
\centering
(b) Constant EoS
\end{minipage}
\label{fig:CWctt}
\caption{\label{fig:contours}Credible intervals of the different observational data sets: \textit{green} ones correspond to GRBs, \textit{purple} to Hubble parameter data, \textit{orange} to SNeIa and \textit{blue} to the combination
of all observational data. To construct these contour plots we have considered a prior on $\Omega_m$ and $w_0$ from WMAP-5years.}
\end{figure*}

Moreover, the present paper innovates in another direction: we choose a combination of datasets with interesting characteristics: we combine the statistically most powerful dataset available, the luminosity measurements of SNeIa, with other datasets:  the luminosity measurements of Gamma Ray Bursts (GRBs) and direct Hubble data. GRBs are particularly useful for the study of oscillating models, as they typically inform us of higher redshifts than supernovae data, they improve the capability for detecting oscillating features (if they exist) at lower frequencies. In addition, the inclusion  of the direct Hubble data can in principle enhance the sensitivity of our tests to the presence of oscillations, as the use of these data does not involve performing an averaging of the inverse of the Hubble factor, and then a possible smoothing of the oscillations is partially compensated for. Another point in favour of the usage of this particular combination of three datasets is the rather good concordance among them (see Fig. \ref{fig:contours}), which applies at least
for the case of a constant $w$ quintessence, and therefore seems a priori a property that will be shared by models with a dynamical
EoS parameter. 

\section{Oscillating dark energy}
\subsection{Our parametrizations}
We have just presented motivations for studying oscillating dark energy. In general,
in order to infer conclusions about the dynamical behaviour of dark energy and its consequences in
the expansion of the Universe, one has to make some concessions to try and make the best
out of a collection of noisy scattered data. One of the most popular is to propose a parametric
reconstruction obeying some basic requirements. Our proposal fits precisely in this kind of approach, and
in order to examine the adequacy of oscillating patterns in the dark energy EoS
we consider a simple and periodic phenomenological parametrization for the
EoS, as proposed and studied in
\cite{Linder06}:
\begin{equation}\label{eq:modelsine}
w(a) = w_{c}-A\sin(B\ln a -\Theta).
\end{equation}
Equation
(\ref{eq:modelsine}) describes evolutionary  dark energy with
\begin{itemize}
  \item $\Theta$ being the phase of the oscillation, which for simplicity is assumed to be
        zero.
  \item $w_c$ being the centre of the range over which $w(a)$
        oscillates. In the   $\Theta=0$ case the parameter $w_c$  is equal to the present value of the dark
        energy EoS, $w(a=1)=w_c=w_0$.
  \item $A$ being the amplitude of the oscillations, which obviously must be non zero for a dynamical $w$.
  If this EoS is the effective realisation of a canonical scalar field, $A$ should fulfil the constrain
        $w_0-A\geq -1$ ; but we leave it completely free and let the observational data speak out its
        value.
  \item $B$ is the frequency of the oscillations; and  there are some key remarks to be made about it.
    In order to notice distinctly the presence of  an oscillatory behaviour
    $B$ should fulfil the condition
        \be \arrowvert B \ln{a_{min}}\arrowvert>2\pi, \ee where $a_{min}$
is given by the highest redshift in the dataset. 
For our observational data this value is associated with a GRB at $z=5.6$, for which
    we infer the constraint $B>3.3$. Nevertheless, as SNeIa represent by far the dataset with the largest statistical power
    in our analysis and they span a smaller redshift range, it seems reasonable that an oscillating pattern will be 
    only detectable for larger values of B.
\end{itemize}

Previous works have attempted to set constraints in these
parameters but they have not been able to distinguish an
oscillatory model from one with a constant EoS. This is partly due
to the data sets employed, but also because of excessive restrictions on the parameters.

In this new attempt at exploring oscillating dark energy we use a new combination
of data sets so as  to obtain more reliable constraints, but we also make a few other key changes: specifically
we carry out an analysis for a discrete set of fixed  $B$ values to try and avoid
the arbitrariness in the choice of this parameter present in works by other authors.

In addition, and given the lack of grounds for very strong restrictions on phenomenological
parametrizations of dark energy we have analysed two more models which represent a slight
departure from the perhaps excessively nicely shaped trigonometric parametrization by Linder.
Our first proposal in this direction is
\begin{equation}\label{eq:modelbessel}
w(a) = w_{c}-A \ J_{1}(B\ln a -\Theta),
\end{equation}
where again we have set the phase $\Theta = 0$ so that $w_{c} =
w_{0}$, and $J_{n}$
is the Bessel function of the first kind with $n = 1$. The Bessel
function $J_{1}(x)$ shows oscillations which are damped with
growing $x$, as opposed to the constant amplitude of the
trigonometric case. So in this case the EoS parameter $w$ would
have an oscillating trend modulated by a damping effect which
makes the amplitude smaller and smaller as we go back in time (in
the redshift space).

As we have to set observational constraints on parameters, we have to compute the Hubble factor
\begin{equation}
H^2(a)=\Omega_ma^{-3}+(1-\Omega_m)a^{-3(1+\overline{w}(a))},
\label{fried}
\end{equation}
which depends on the present value of the fractional matter density $\Omega_m$, and on the averaged equation of state
parameter
\begin{equation}
 \overline{w}(a)=\frac{\int_0^{\ln(a)}w(a)d\ln(a)}{\ln(a)}.
\label{eq:eqofstateav}
\end{equation}

So, we will work at some points with this expression of the EoS parameter 
instead of the usual one.
It can be noticed that for the  two parametrizations presented above, and we
 realised that the averaged forms follow a common pattern; the parametrization
given by the Eq. \ref{eq:modelsine} takes the form
\begin{equation}
\overline{w}(a) = w_{0}+\frac{A}{B}\frac{\cos(B\ln a)-1}{\ln(a)},
\label{eq:modelsineav}
\end{equation}
whereas the parametrization given by Eq.\ref{eq:modelbessel}
\begin{equation}
\overline{w}(a) = w_{0}+\frac{A}{B}\frac{J_0(B\ln a)-1}{\ln(a)} .
\label{eq:modelbesselav}
\end{equation}
With  Eq. \ref{eq:modelbesselav} and Eq. \ref{eq:modelsineav} as a 
reference, we propose another parametrization with a damped oscillating behaviour,
\begin{equation}
 \overline{w}(a) = w_{0}+\frac{A}{B}\frac{(\pi/2)H_{-1}(B\ln a)-1}{\ln(a)}
\end{equation}
where $w_0$ is the value of the EoS at present and $H_{-1}$ is the Struve function,
$H_\alpha$, of order $\alpha=-1$. The Struve function lies between
the trigonometric case, with constant amplitude, and the Bessel
function, being less acutely damped than the Bessel function.
It can be checked that the $w(a)$ function for this new proposal
is an oscillating one as well, so this new model fits is the aim
of the discussion.

Anyway, we have to underline that our three $w$ models depend on $\ln a$
(i.e. $\ln (1+z)$) so these effects are quite smoothed and not so
evident when depicted on even a large redshift interval such as the one
limited by the re-ionisation one ($z = 1089$). However, such parametrizations, as
first shown by \cite{Linder06}, allow for analytical expressions of $H(z)$ 
(see Eq. \ref{eq:eqofstateav}) and thus
result convenient for evaluation purposes.

Interestingly, as we will discuss later, the main consequences of an oscillating equation of
state are more evident on the deceleration $q(z)$ function than on the Hubble function or the dark energy fractional
density $\Omega_{X}$.

\subsection{Earlier Works}
\label{sec:Oscmodel_past}

In \cite{Riess07} an interesting observational clue of a possible
oscillating behaviour in the EoS was found. Fitting a quartic polynomial of $w(z)$  to SNeIa observations indications that such oscillations might be really present were found. Even though the redshift range where
these oscillations seem to be present is actually rather small;
these results motivated a plethora of works trying to
analyse this possibility deeper.

In \cite{Xia05} the formulation
\begin{equation}
w(z) = w_{0} + w_{1} \sin(z),
\end{equation}
was used to  model a quintom scenario   phenomenologically. The analysis was conducted using
luminosity distances of SNeIa, the CMB shift parameter and the
linear growth rate from large scale structure. Combining all these
data sets the best fit values found were $w_{0} = -1.33$, $w_{1} = 1.47$. Such oscillating model
differs very little from the linear one ($w(z) = w_{0} + w_{1} z$) when considering constraints from SNeIa only, and the oscillating case is only mildly preferred. This was somehow to be expected as for low redshifts the two parametrizations are very similar.  Note that as in this study the oscillation frequency was fixed to a specific value we meet again the difficulties to make general inferences.

An extension of the latter was done in \cite{Xia06}, where
\begin{equation}
w(a)=w_0+w_1\sin\left(w_2\ln a +w_3\right),
\end{equation}
was proposed,  therefore now including also a period and a phase. One further degree of sophistication
was introduced in the analysis as it was carried out using a modified version of CosmoMC so as
to have into account dark energy perturbations.
The results indicate that $w_{0}$ is acceptably well constrained, while
$w_{1}$ is poorly constrained, and  $w_{2}$ and $w_{3}$ can be regarded  as completely unconstrained.
This is not surprising, though, as highly dimensional parametrizations of dark energy seem to suffer this problem
typically given the precision of observational data at present.

In \cite{Xia08}  constraints on the cosmological
parameters of some oscillating models were analysed using simulated data from future Planck
measurements. Two different parametrizations of the dark
energy EoS were used, the CPL one
\begin{equation}
w(a)=w_0+w_{1}\left(1-a\right)
\end{equation}
and the a reformulation of the \cite{Linder06} oscillating one,
\begin{equation}
w(a)=w_0+w_{1}\sin(w_{2}\ln{\left(a\right)}),
\end{equation}
having fixed the phase parameter to zero. In addition, one
further choice was made: for simplicity and for focusing the analysis on
low redshifts,  the period parameter was fixed as $w_{2} =
3\pi/2$ (because it corresponds to a period in the redshift range
$0<z<2$ where data from supernovae are more robust). Then they use
present data from WMAP3, ESSENCE SNeIa, SDSS (for details see
reference) to derive best fit values for the EoS parameters which are
assumed as the fiducial values for deriving constraints from
Planck mission details. The following results were obtained:
\begin{equation}
w_0 =-1.03 ^{+0.15+0.36}_{-0.15-0.26}, \quad w_1 =-1.03
^{+0.562+0.781}_{-0.587-1.570}
\end{equation}
for the CPL EoS, and
\begin{equation}
w_0 =-0.981 ^{+0.320+0.534}_{-0.340-0.748}, \quad w_1 =-0.068
^{+0.561+1.037}_{-0.591-1.245}
\end{equation}
for the oscillating model.

Assuming the CPL EoS they are able to conclude that a Quintom scenario
with a $w(z)$ which crosses the phantom divide is
favoured against the $\Lambda$CDM scenario (it is mildly favoured with
present data while future surveys can provide narrower constrains).
Assuming the oscillating EoS the
same conclusion can be derived, even if the constraints are weaker, and the low value for the amplitude of oscillation
makes the constant $w(z)$ (i.e. $\Lambda$CDM model) equally possible.

For what concerns the oscillating model, these results are backed
up in \cite{Liu09}, where the following $68\%$ CL values were
obtained:
\begin{eqnarray}
w_0=-0.958\pm 0.098, \quad w_{amp} =0.030 ^{+0.124}_{-0.130}
\end{eqnarray}
From these results one can conclude that the EoS fit with current
data provides as good as $\Lambda$CDM,  and that the oscillation amplitude  is
limited: $\arrowvert w_{amp}\arrowvert <0.232$ at $95\%$CL.

An oscillating model different from the one proposed by Linder is
analysed in \cite{Jan07} (inspired by a previous idea in \cite{Feng06}).
The authors  used SNeIa, the CMB shift parameter and the measurement of
the BAO peak from SDSS in order to constrain the sine version of the
oscillating EoS written in the form
\begin{equation}
w(a) = w_{0}-A\sin(B\ln a)
\end{equation}
and the alternative
\begin{equation}
w(a) = -\cos(b\ln a).
\end{equation}
The period parameter $b$ in the second model drives the accelerating/decelerating
epochs, and in the limit of small $b$ one has $w(a) \sim -1$.

Final results show that $b = 0.06 \pm 0.01$ at $1\sigma$ level, a
value which is consistent with limits required by CMB and correct
power spectrum and which represents a \textit{clear} evidence for an
oscillating behaviour with a very long period (i.e. small
but clear deviation from cosmological constant).

Another interesting work is \cite{Kurek08}, where  many different  EoS
were compared with a Bayesian approach. They worked with the so
called linear EoS, i.e. CPL model; with pure oscillating models,
the cosine one, $w(z) = w_{0} \cos (w_{c} \ln (1+z))$, and the
sine one, $w(z) = -1 + w_{0} \sin (w_{s} \ln (1+z))$; with a
damped version of previous oscillating models, $w(z) = w_{0}
(1+z)^3 \cos (w_{c} \ln (1+z))$ or $w(z) = -1 + w_{0} (1+z)^3 \sin
(w_{s} \ln (1+z))$; and with a more complicated version of the
dark energy EoS directly derived from the dynamics of the phantom
scalar field (see reference for details). Analysing the values
of the Bayesian evidences obtained fitting data (SNeIa, CMB, BAO)
for each particular model, and comparing them with the Jeffreys
scale (\cite{Jeffreys}), they concluded that there is a
substantial evidence for preferring pure oscillating model (sine)
and the one derived from phantom field dynamics over the linear
one, while damped versions are completely excluded. In addition,
they found  no strong evidence to favour $\Lambda$CDM over the
oscillating models.

As stated in the Introduction, completely different approaches
are possible as well; for example, in \cite{Saez09} the starting point was an
oscillating model for $H(z)$ instead of $w(z)$. Finally, it is
possible to consider oscillating fields, which do  not necessarily
produce an oscillating EoS \cite{Frieman95}.

\section{Observational data}
\label{sec:Obstest}

We have tested the possible periodicity of the Hubble function by using three
different observational data sets, i.e.:
\begin{itemize}
   \item the reconstructed Hubble data given in \cite{Stern09};
   \item the \textit{Constitution} Supernovae data set described in
         \cite{Hicken09};
   \item the Gamma Ray bursts luminosity distances measured and analysed in
         \cite{Kodama08}.
\end{itemize}

\subsection{Hubble parameter: Stern et al. 2009 data set}
\label{sec:Hubbledata}

Recently in \cite{Stern09} an update of the Hubble function $H(z)$ data extracted
from differential ages of passively evolving galaxies previously
published in \cite{Simon05} was presented. Constraining the
background evolution of the Universe using these data is
interesting for several reasons. Firstly, they can be used together with
other cosmological tests in order to get useful consistency checks
or tighter constraints on models. Secondly, and more 
importantly, in contrast to standard candle luminosity distances or
standard ruler angular diameter distances, the Hubble function is
not integrated over. This is a key point because if a periodic
behaviour is present in $w(a)$ it should be directly detectable in
$H(a)$ while it could be lost in luminosity or angular diameter
distances because of integration stages.

The Hubble parameter dependence on the differential age of the
Universe in terms of redshift is given by
\begin{equation}
H(z)=-\frac{1}{1+z}\frac{dz}{dt}.
\end{equation}
Thus, $H(z)$ can be determined from measurements of $dt/dz$. As
reported in \cite{Jimenez02}, \cite{Jimenez03}, \cite{Simon05} and
\cite{Stern09}, values of $dt/dz$ can be computed using absolute
ages of passively evolving galaxies.

The galaxy spectral data used by \cite{Stern09} come from
observations of bright cluster galaxies done with the Keck/LRIS
instrument\footnote{See \cite{Stern09B} for a detailed description
of the observations, reductions and the catalog of all the
measured redshifts}. The purposely planned Keck-survey
observations have been extended with other datasets: SDSS
improvements in calibration available in the Public Data Release 6
(DR6) have been applied to data in \cite{Jimenez03}; the SPICES
infrared-selected galaxies sample in \cite{Stern00}; and the VVDIS
survey by the VLT/ESO telescope in \cite{LeFevre05}.

The authors of these references bin together galaxies with a
redshift separation which is small enough so that the galaxies in
the bin have roughly the same age; then, they calculate age
differences between bins which have a small age difference which
is at the same time larger than the error in the age itself
(\cite{Stern09}). The outcome of this process is a set of 11 values
of the Hubble parameter versus redshift. A particularly nice
feature  of this test is that the sensitivity of differential ages
to systematic error is lower than in the case of absolute ages
(\cite{Jimenez04}).

Observed values of $H(z)$ can be used to estimate DE parameters by
minimising the quantity
\begin{equation}\label{eq: Hub_chi}
\chi^{2}_{\mathrm{H}}(H_{0}, \{\theta_{i}\}) = \sum^{9}_{j = 1}
\frac{(H(z_{j}; \{\theta_{i})\} -
H_{obs}(z_{j}))^{2}}{\sigma^{2}_{\mathrm{H}}(z_{j})}
\end{equation}
where $H_{0} \doteq 100 \, h$ will be fixed as $h= 0.742$ (\cite{Riess09}), while
the vector of model parameters, $\theta_{i}$, will be
$\theta_{i}=(\Omega_{m}, w_{0}, A, B)$.

\subsection{Supernovae: Hicken et al. 2009 data set}
\label{sec:SNdata}

We use one of the most recent SNeIa samples, the
\textit{Constitution} sample described in \cite{Hicken09}, which is a 
data set obtained by combining
the Union data set by \cite{Kowalski08} with new $90$ nearby
objects from the CfA3 release described in \cite{Hicken09A}.

The Union SNe compilation is a data set of low-redshift
nearby-Hubble-flow SNe and is built with new analysis procedures
for working with several heterogeneous SNeIa compilations. It
includes $13$ independent sets with SNe from the SCP, High-z
Supernovae Search (HZSNS) team, Supernovae Legacy Survey and
ESSENCE Survey, the older data sets, as well as the recently
extended data set of distant supernovae observed with HST. After
various selection cuts were applied in order to create a
homogeneous and high-signal-to-noise data set, a final collection of $307$
SNeIa events distributed over the redshift interval $0.15 \leq z
\leq 1.55$ was obtained.

The CfA3 data set was  originally made of 185 multi-band optical
SNeIa light curves obtained at the F.L. Whipple Observatory of the
Harvard-Smithsonian Center for Astrophysics (CfA); 90 of the
original 185 objects passed the quality cuts of \cite{Kowalski08}
and were added to the Union data set to form the Constitution
one.

The statistical analysis of the Constitution SNe sample rests on the
definition of the modulus distance,
\begin{equation}
\mu(z_{j}) = 5 \log_{10} [ d_{L}(z_{j}, \{\theta_{i}\}) ]+\mu_0,
\end{equation}
where $d_{L}(z_{j}, \{\theta_{i}\})$ is the Hubble free luminosity
distance
\begin{equation}\label{eq:dl_H}
d_{L}(z,\{\theta_{i}\}) = (1+z) \ \int_{0}^{z} \mathrm{d}z'
\frac{1}{H(z',\{\theta_{i}\})}.
\end{equation}
The best fits to be presented will be obtained by minimising the quantity
\begin{equation}\label{eq: sn_chi}
\chi^{2}_{\mathrm{SN}}(\mu_{0}, \{\theta_{i}\}) = \sum^{397}_{j =
1} \frac{(\mu(z_{j}; \mu_{0}, \{\theta_{i})\} -
\mu_{obs}(z_{j}))^{2}}{\sigma^{2}_{\mathrm{\mu}}(z_{j})}
\end{equation}
where the $\sigma^{2}_{\mathrm{\mu}}$ are the measurement
variances. The nuisance parameter $\mu_{0}$ encodes the Hubble
parameter and the absolute magnitude $M$, and has to be
marginalised over. Giving the heterogeneous origin of the Constitution data
set, and the procedures described in \cite{Kowalski08} and
\cite{Hicken09} for reducing data, we have worked with an
alternative version of Eq.~(\ref{eq: sn_chi}), which consists in
minimizing the quantity
\begin{equation}\label{eq: sn_chi_mod}
\tilde{\chi}^{2}_{\mathrm{SN}}(\{\theta_{i}\}) = c_{1} -
\frac{c^{2}_{2}}{c_{3}}
\end{equation}
with respect to the other parameters. Here
\begin{equation}
c_{1} = \sum^{307}_{j = 1} \frac{(\mu(z_{j}; \mu_{0}=0,
\{\theta_{i})\} -
\mu_{obs}(z_{j}))^{2}}{\sigma^{2}_{\mathrm{\mu}}(z_{j})}\, ,
\end{equation}
\begin{equation}
c_{2} = \sum^{307}_{j = 1} \frac{(\mu(z_{j}; \mu_{0}=0,
\{\theta_{i})\} -
\mu_{obs}(z_{j}))}{\sigma^{2}_{\mathrm{\mu}}(z_{j})}\, ,
\end{equation}
\begin{equation}
c_{3} = \sum^{307}_{j = 1}
\frac{1}{\sigma^{2}_{\mathrm{\mu}}(z_{j})}\,.
\end{equation}
It is trivial to see that $\tilde{\chi}^{2}_{SN}$ is just a
version of $\chi^{2}_{SN}$, minimised with respect to $\mu_{0}$.
To that end it suffices to notice that
\begin{equation}
\chi^{2}_{\mathrm{SN}}(\mu_{0}, \{\theta_{i}\}) = c_{1} - 2 c_{2}
\mu_{0} + c_{3} \mu^{2}_{0} \,
\end{equation}
which clearly becomes minimum for $\mu_{0} = c_{2}/c_{3}$, and so
we can see $\tilde{\chi}^{2}_{\mathrm{SN}} \equiv
\chi^{2}_{\mathrm{SN}}(\mu_{0} = 0, \{\theta_{i}\})$. Furthermore,
one can check that the difference between $\chi^{2}_{SN}$ and
$\tilde{\chi}^{2}_{SN}$ is negligible.

\subsection{GRBs: Kodama et al. 2008 data set}
\label{sec:GRBdata}

The GRBs sample, described in \cite{Kodama08}, is
made of 33 GRBs within the redshift interval $z<1.62$ and 30 GRBs
in the redshift interval $1.8 < z < 5.6$. It is well known that
GRBs are not standard candles as SNeIa; at the same time they
contain a lot of important information about high redshift properties
of the universe which cannot be derived from SNeIa. So their combined
use can bring important and complementary information about the
reconstruction of dark energy and gives us the possibility to
detect eventually traces of an oscillatory behaviour on a larger
redshift range. The calibration of GRBs data can be done in
several ways, and many empirical formulas have been given for
describing the peak energy-peak luminosity correlation. In
\cite{Kodama08} the peak energy-peak luminosity correlation
is described by the so called \textit{Yonetoku relation}\footnote{See
\cite{Yonetoku04}} of the GRBs sub-sample in redshift interval
$z<1.62$ is calibrated without assuming any cosmological model and
using the luminosity distance of the objects considered it is estimated from
those of SNeIa with redshift $z<1.755$. The calibrated Yonetoku
relation is then finally applied to the high redshift GRBs
sub-sample with redshift in the interval $1.8 < z < 5.6$. Final
data consist of a set of calibrated luminosity distances, so that
we can define the contribution to the total chi-square as:
\begin{equation}\label{eq:chi_grbs}
\chi^{2}_{\mathrm{GRB}}(\{\theta_{i}\}) = \sum^{63}_{j = 1}
\frac{(d_{L}(z_{j}; \{\theta_{i})\} -
d_{L}^{obs}(z_{j}))^{2}}{\sigma^{2}_{\mathrm{d_{L}}}(z_{j})}
\end{equation}
where the $\sigma^{2}_{\mathrm{d_{L}}}$ are the measurement
variances.

\section{Statistics and data analysis}
\label{sec:Markov}

We will explore the probability distributions of our problem with Markov Chain Monte Carlo (MCMC) methods.
These methods (fully described in \cite{Berg}, \cite{MacKay},
\cite{Neal} and references therein) extract samples sequentially
using a probabilistic algorithm (we have chosen the
Metropolis-Hastings algorithm which we describe below) based on
the Bayesian statistical approach. The main problem when running a
Markov chain concerns how to avoid biased inferences or
underestimations on errors of the theoretical parameters. The
\textit{problem of the convergence} is the main one among the
problems of that sort, and it can be formulated as follows: how
can one be sure that the properties of a sample from the MCMC
algorithm are a good representation of the unknown distribution to
be explored? In the following we discuss the details
algorithm and the solution adopted for convergence.\\

\subsection{Markov Chain algorithm and convergence test}
\label{sec:MCMC}

The Metropolis-Hastings algorithm works as follows: starting from
an initial parameter vector $\mathbf{p}$ (in our case $\mathbf{p}
= (\Omega_{m}, w_{0}, A, B)$),  one generates a new trial point
$\mathbf{p'}$ from a \textit{proposal density}
$q(\mathbf{p'},\mathbf{p})$, which represents the conditional
probability to have $\mathbf{p'}$ given $\mathbf{p}$. This new
point is accepted with probability
\begin{equation}
\alpha(\mathbf{p}, \mathbf{p'}) = {\rm min} \left\{1,
\frac{P(\mathbf{p'}|\mathbf{d})
q(\mathbf{p'},\mathbf{p})}{P(\mathbf{p}|\mathbf{d})
q(\mathbf{p},\mathbf{p'})}\ \right\}
\end{equation}
where $P(\mathbf{p}|\mathbf{d})$ is the conditional probability to
have the parameter set $\mathbf{p}$ given the  observational data
$\mathbf{d}$. Then, from Bayes' theorem it follows that this
probability is:
\begin{equation}
P(\mathbf{p}|\mathbf{d}) \propto L(\mathbf{d}|\mathbf{p})
P(\mathbf{p})
\end{equation}
with $L(\mathbf{d}|\mathbf{p'})$ being the likelihood function
containing information from the data, and $P(\mathbf{p})$ the
prior on the parameters which contains all the supposed information
on them before observing the data. If the chain moves to the new set
$\mathbf{p'}$, then one says that it has been accepted, otherwise it
has been rejected. 

We choose the prior depending on the physical
requirements a given parameter has. The proposal density is typically a Gaussian
symmetric with respect to the two vectors $\mathbf{p}$ and
$\mathbf{p'}$, namely $q(\mathbf{p},\mathbf{p'}) \propto
\exp(-\Delta p^{2} / 2 \sigma_T^{2})$, with $\Delta \mathbf{p} =
\mathbf{p} - \mathbf{p'}$, so that, from the detailed balance
equation, we know that the final probability distribution is
stationary under the Markov process. An important quantity in the
testing proposal distribution is its dispersion $\sigma_{T}$; as
we will discuss below, we decide not to take a fixed value for it,
but instead, we let  it depend on the value of the parameters at any
step.

One can say that a chain has reached convergence when the
statistical properties of the extracted samples can describe the
statistical properties of the unknown probability distribution
with \textit{good accuracy}. Probability theory says that Markov
processes will reach the exact final distribution in an asymptotic
way, which means a sample's length will become infinite in an
infinite computation time. Of course, as one is forced to operate with finite
length samples, the question arises of how this truncation
can bias the final statistical results, and if there are any
parameters that can be able to give information about the goodness
of the process. In the literature the most used parameter for this
task is the \textit{convergence ratio} (see \cite{Dunkley05}),
defined as
\begin{equation}
r = \frac{\sigma_{\overline{x}}^{2}}{\sigma_{0}^{2}}.
\end{equation}
This is the ratio between the variance of the mean of the samples
and the variance of the underlying unknown distribution (we will
operate with standard distributions so that $\sigma_{0}^{2}=1$).
Then $r$ is required to be below a cut-off limit value, typically
$0.01$, to have a guaranteed convergence of the chain. This
parameter is used in  other convergence tests, such as the
Gelmann-Rubin test (see \cite{GelmanRubin}), which runs many
parallel multiple chains and estimates $r$ at any step; but we
regard it a time and hardware expensive test.

An alternative solution to this problem in the spectral analysis
approach proposed by \cite{Dunkley05}. Is it clear that all the
steps in MCMC are correlated; this correlation is somewhat
intrinsic to the code, at least before having reached convergence,
but it depends also on the value of $\sigma_{T}$\footnote{For a
detailed analysis of this aspect see \cite{Dunkley05}}. But what
is even more important is the behaviour of the correlation when
the convergence has been reached: in this case the MCMC will
sample from the underlying distribution and it will work like a
random sampler, so that there will be no correlations in this regime.

This is the key idea of the test by \cite{Dunkley05}: if we take
the \textit{power spectra} of the MCMC samples, we will have a
large correlation on small scales, but the spectrum will become flat
(like a white noise spectrum) when convergence has been reached.
Then, checking the spectrum of just one chain (instead of many
parallel chains as in Gelmann-Rubin's test) will be sufficient to
assess that convergence has indeed been reached. We will give just a
short account of the steps to be followed to implement the test,
but for  a more detailed reference see \cite{Dunkley05}.

In brief, we calculate the discrete power spectrum of the chains,
\begin{equation}
P_{j} = |a_{N}^{j}|^{2},
\end{equation}
with
\begin{equation}
a_{N}^{j} = \frac{1}{\sqrt{N}}\sum_{n=0}^{N-1} x_{n}
\exp{\left[i\frac{2\pi j}{N}n\right]},
\end{equation}
where $N$ and $x_{n}$ are respectively the length and a given
element of the sample from the MCMC, $j=1,\ldots,{N}/{2}-1$, and
the wave number $k_{j}$ of the spectrum is related to the index $j$
by the relation $k_{j}={2\pi j}/{N}$. Then we fit it to an
analytical template:
\begin{equation}
P(k) = P_{0} \frac{(k^{*}/k)^{\alpha}}{1+(k^{*}/k)^{\alpha}},
\end{equation}
or in equivalent logarithmic form:
\begin{equation}
\ln P_{j} = \ln P_{0} + \ln
\left[\frac{(k^{*}/k_{j})^{\alpha}}{1+(k^{*}/k_{j})^{\alpha}}\right]-\gamma
+ r_{j},
\end{equation}
where $\gamma=0.57216$ is the Euler-Mascheroni number and $r_{j}$
are random measurement errors with $<r_{j}> = 0$ and $<r_{i}r_{j}>
= \delta_{ij} \pi^{2}/6$. The fit provides  estimates of three
parameters, but only two of them are fundamental to our analysis.
The first one is $P_{0}$, which is the value of the power spectrum
extrapolated for $k \rightarrow 0$; this is an important parameter
because from it we can derive the convergence ratio using $r
\approx {P_{0}}/{N}$. The second important parameter is $j^{*}$
(the index corresponding to $k^{*}$), which is related to the
turnover point from a power to a flat spectrum; the estimated
value of $j^{*}$ has to be $\gtrsim>20$, so one can be sure that the
number of points in the sample coming from the convergence region
are larger than the number of noisy points. If these two
conditions are met for all the parameters, then the chain has
reached convergence, and the statistics from the MCMC procedure describes well
the underlying probability distribution. Following the  advise in
\cite{Dunkley05} we perform the fit over the range $1 \leq j \leq
j_{max}$, with $j_{max} \sim 10 j^{*}$, where a first estimation
of $j^{*}$ can be obtained from a fit with $j_{max} = 1000$, and
then perform a second (or even a third) iteration to have a better
estimation of it.

\subsection{Model selection tests}
\label{sec:testmodel}

After having estimated the value of the parameters set by using the
MCMC approach, we need a tool for comparing, selecting and testing the
statistical goodness of our results. The related literature is too
extensive for being reviewed here, so we will only refer in some detail to  the tools we have chosen.

Since the MCMC technique is based on a Bayesian approach, the best way for
comparing models is arguably the Bayesian Evidence. It
is defined as
\begin{equation}
E \equiv \int \mathcal{L}(\mathbf{\theta}) P(\mathbf{\theta})
\mathrm{d}\mathbf{\theta} \, ,
\end{equation}
where $\mathcal{L}$ is the likelihood function, $\mathbf{\theta}$
is the parameters vector and $P(\mathbf{\theta})$ is the prior
distribution for the parameters. It is clear from its definition,
that the evidence of a model is the average likelihood of the
model with respect to the prior: models which fit the data well
and make narrow predictions are likely to fit well over much of
their available parameter space, giving a high evidence. So using
evidence for comparing models is a very appropriate task. Model comparison requires defining then 
Bayes factor,
\begin{equation}
B_{ij} = \frac{E(M_{i})}{E(M_{j})},
\end{equation}
which is the ratio between the evidence values of two models, $M_{i}$
and $M_{j}$. If $B_{ij}>1$ then the model $M_{i}$ is preferred
with respect to the model $M_{j}$. By convention, Bayes factor is
judged on the Jeffreys' scale (\cite{Jeffreys}): for $1<\ln
B_{ij}<2.5$ there is a ``substantial'' evidence in favour of the
model with the greatest Bayesian evidence; for $\ln B_{ij}>5$ the
evidence is ``decisive''.

Moreover,  we have to underline in favour of the Bayesian evidence,
that it is a full implementation of Bayesian inference and can be
directly interpreted in terms of model probabilities.
Unfortunately, being a highly-peaked multi-dimensional integral,
its estimation typically requires a hard and challenging numerical effort.
Even if some algorithms have been found which simplify this
operation, it is always preferable to have easier tools for
estimating it and comparing models.

The easiest and most used tools are different versions of the
Information Criteria. Generally, the introduction of a higher
number of parameters improves the fit to the chosen dataset,
regardless of whether or not these new parameters are really
relevant. As a consequence, the simple comparison of the
maximum likelihood value of different models will tend to favour
the model with the highest number of parameters. The information
criteria work just in this direction: they compensate this
behaviour by penalising models which have more parameters.

The first test is the Akaike Information Criterion (AIC) defined
as
\begin{equation}
\mathrm{AIC} = -2 \ln \mathcal{L} + 2 k
\end{equation}
where $\mathcal{L}$ is the maximum likelihood value and $k$ is the
number of parameters of the model (\cite{Akaike74}). The
$\mathrm{AIC}$ is derived by an approximation of the
Kullback-Leibler information entropy, which measures the
difference between the true data distribution and the model one
(\cite{Burnham02} and \cite{Takeuchi00}). The best model is the
one which minimises the $\mathrm{AIC}$, and no requirement for the
models is asked for.

There also exists an $\mathrm{AIC}$ version for small sample
sizes, the corrected $\mathrm{AIC}$, $\mathrm{AIC_{c}}$
(\cite{Burnham02}) given by
\begin{equation}
\mathrm{AIC_{c}} = \mathrm{AIC} + \frac{2k(k+1)}{N-k-1},
\end{equation}
where $N$ is the number of points in the dataset. Since the
correction term disappears for large sample sizes, $N >> k$, we
will use this last definition for comparing models (as 
pointed out in \cite{Liddle07}, it is always preferable to use the
corrected version rather than the original one).

From the same principles of AIC (minimisation of the
Kullback-Leibler information entropy) another comparing
tool can be derived, the Residual Information Criterion (RIC). We will use the
corrected version in \cite{Leng07}, where the $\mathrm{RIC_{c}}$
is defined as
\begin{equation}
\mathrm{RIC_{c}} = -2 \ln \mathcal{L} + (k-1) + \frac{4k}{N-k-1}
\,.
\end{equation}
When $N
>> k$, $\mathrm{RIC_{c}}$ has a smaller penalty than
$\mathrm{AIC_{c}}$.

Generally, when $\Delta \mathrm{AIC} \gtrsim 1$,  it follows
that the two
models are significantly different, and the one with the
lowest value of AICid the preferred one. Finally, we have to keep in mind
that AIC tends to favour models with a high number of parameters and it is
``dimensionally inconsistent'', namely, that even as the dataset
size tends to infinity, the probability of the AIC incorrectly
picking an overparametrized model does not tend to zero
(\cite{Liddle04} and references therein). Let us thus consider alternatives.

The Bayesian Information criterion ($\mathrm{BIC}$) was introduced
in \cite{Schwarz78} and is defined as
\begin{equation}
\mathrm{BIC} = -2 \ln \mathcal{L} + k \ln N \; .
\end{equation}
Again in this case the best model has the lowest BIC, and it is
clear from this expression that BIC penalises models with a high
number of parameters more than AIC. Being the BIC a good
approximation for twice the log of the Bayes factor, it can be
compared with Jeffreys' scale, so that $\Delta \mathrm{BIC}>5$
means a ``strong'' evidence in favour of the model with lowest BIC
values, while for $\Delta \mathrm{BIC}>10$ this evidence is
``decisive''.

We will also use the Deviance information criterion (DIC) of
\cite{Spiegelhalter02}, which mixes elements from both Bayesian and
information theory. It is well suited to our case
because it is easily computable from posterior samples such as
those coming from MCMC runs. It relies on the definition of the
effective number of parameters, $p_{D}$, also known as the
Bayesian complexity. It is defined as
\begin{equation}
p_{D} = \overline{D(\mathbf{\theta})} -
D(\overline{\mathbf{\theta}}),
\end{equation}
where
\begin{equation}
D(\mathbf{\theta}) = -2 \ln \mathcal{L}(\mathbf{\theta}) + C
\end{equation}
with $C$ a constant which vanishes from any derived quantity, and
the chi-square defined as usual as $\chi^{2} = -2 \ln
\mathcal{L}$. This definition shows that $p_{D}$ can be considered
as the mean deviance minus the deviance of the means, and this is
the key quantity in estimating the degrees of freedom of a test.
Finally the DIC is defined by
\begin{equation}
\mathrm{DIC} = D(\overline{\mathbf{\theta})} + 2 p_{D} =
\overline{D(\mathbf{\theta})} + p_{D}
\end{equation}
where one can recognise a similar-to-AIC formulation in the first
expression, while a Bayesian definition and measure of model
adequacy (penalised by an additional term related to the model
dimensionality) is implicit in the second one. The DIC is also
useful for another reason: it overcomes the difficulties  AIC and BIC have to 
discount parameters which are unconstrained by data (BIC is of
even more suspicious validity when there is any parameter
degeneracy). Finally, $\mathrm{DIC}$ (like $\mathrm{BIC}$) is not
dimensionally inconsistent, so it is able to detect wrong high
dimensionality parametrizations.

\section{Constraints and assumptions}

We have implemented a few priors for running our MCMCs. The main
one has been to set the control $0<\Omega_m<1$, which is a minimum
physicality requirement. We have also set  mild Gaussian priors on $\Omega_m$ and $w0$ with the $3\sigma$ error
bar from WMAP5 as a reference. 

As we stated in the previous sections it would be possible
to set physical limits on the parameters of the oscillating model,
such as $A$ and $B$. The amplitude $A$ should have a value which
depends on the theoretical scenario chosen to be the
fiducial one. In the case of $\Lambda$CDM, one could require that
the minimum value for the EoS was $w_{0} - A \geq -1$. But it is
clear that with this case will exclude the possibility of
a phantom behaviour, which is a possibility present
data do not rule at and in some cases seem to be the preferred one. 
For that reason regard leaving $A$ free as the best option.

At the same time, the frequency (period) parameter $B$ should be suject to $\arrowvert B
\ln{a_{min}}\arrowvert>2\pi$. The highest
redshift of our observational data corresponds to a GRB observation at
$z=5.6$. For that choice, $B$ should be $B>3.3$. Anyway, we have no a priori
strong clues about the validity of a periodic oscillating EoS, so there is scarce
guidance regarding a suitable lower bound on $B$. We could have
oscillations with a period bound given by the highest  redshift in the supernovae data, so 
in this case we would have $B>5.7$; or 
we could have no detectable oscillations at all in our redshift
range so that a really small value of  $B$ could turn out to be the best fit. 
In addition, rigourosly the bound on B we are commenting about (proposed by Linder) makes only sense for the sine
oscillating model, which has a valid definition for the
\textit{oscillation period} and not for other two models.
Anyway, also in this cases the $B$ parameters can be related to
periodic properties of the dark EoS parameter, so we treated it on the same footing
in the analysis of the three models.

There is also another problem concerning the number of free
parameters one introduces in a model. In our case we would have a
dark energy EoS with three free parameters, namely $w_{0}$, $A$
and $B$ (which become four parameters because of the presence of
$\Omega_{m}$ in the expression for the Hubble function and luminosity
distance). This poses a well known problem in reconstructing or
modeling the EoS with parametric relations: how many parameters
can we enquire about?
Following the Occar's razor prescription one could be tempted to
choose the minimalistic option: models with few
parameters are the best ones. But sometimes this could be a not
\textit{physically} good choice: complex systems could require
complex analytical formulas and a large number of parameters could
describe a more suitable behaviour of dark energy. It is also
possible that not all the introduced parameters are really free,
and there is a correlation/dependence between some of them; but
this cannot be known a priori when proposing a new model.

The only solution is to decide depending on the physical problem
one has to face with. In preliminary runs  we left 
all the parameters free, but it soon emerged that there is
a strong degeneracy between some of them, mainly between $A$ and $B$.
While $\Omega_{m}$ and $w_{0}$ were well constrained, the two
main parameters of our oscillating models showed a degeneracy which
made them eventually unconstrained and yielded no satisfactory
information about our proposed EoS.

So we turned to another way to proceed: we fixed
the value of $B$ to a set of discrete values scanning entirely
the range of  values which could potentially lead to  detectable
oscillations of our observable functions given the criteria discussed above.

\section{Results}

\begin{figure}
\begin{center}
\includegraphics[width=0.35\textwidth]{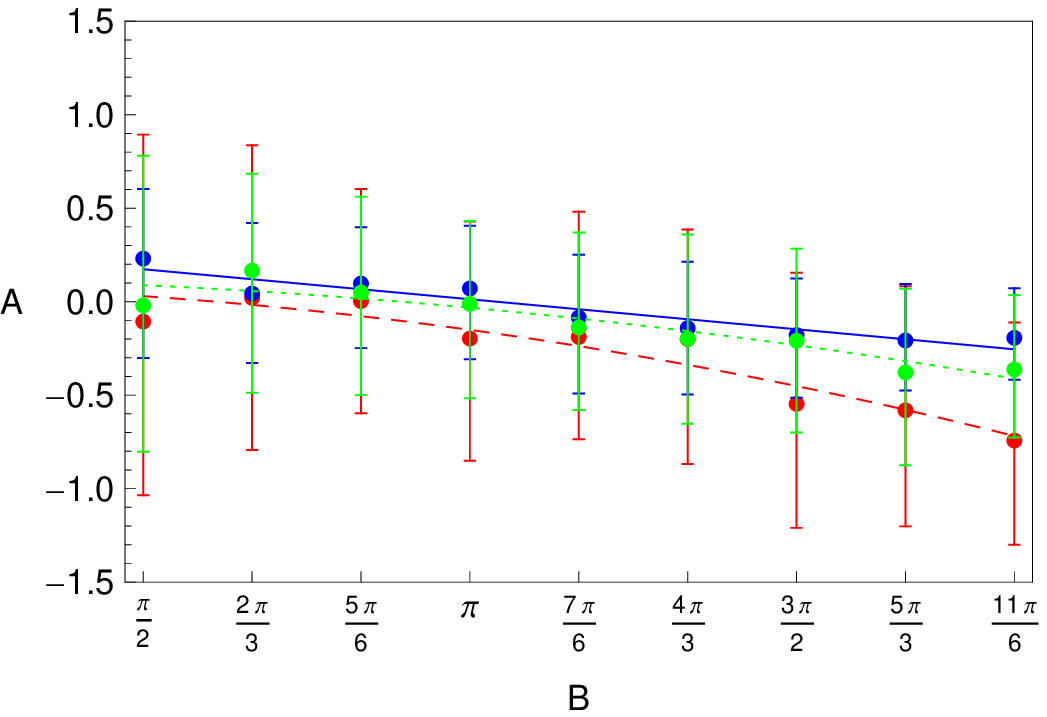}
\label{fig:AvsBcom}
\includegraphics[width=0.35\textwidth]{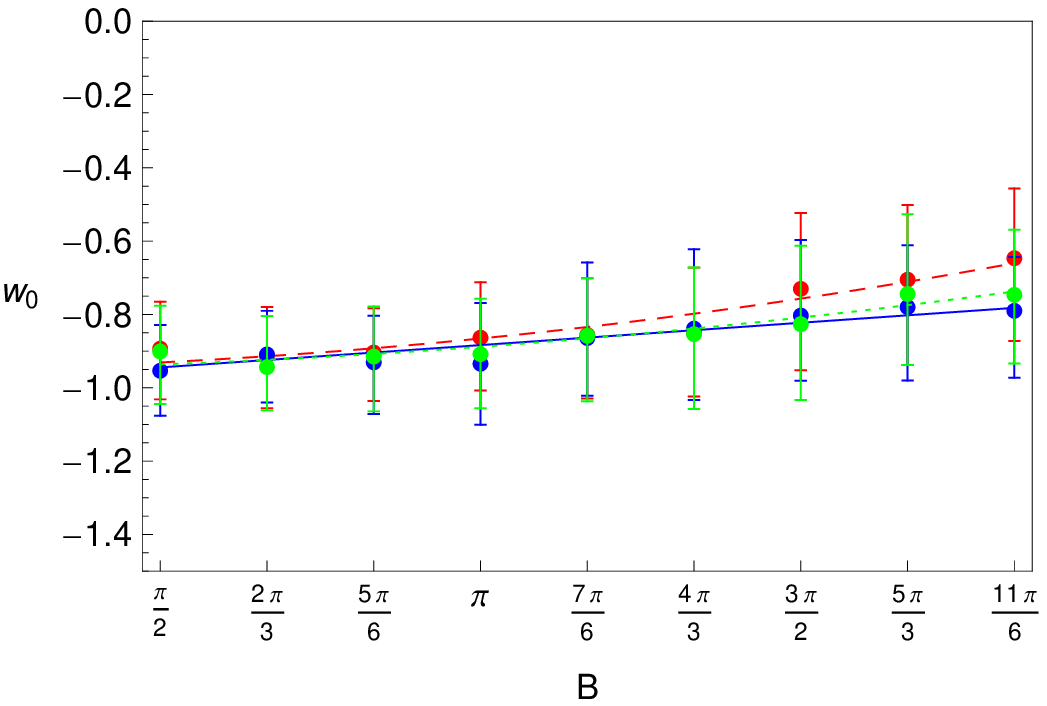}
\label{fig:w0vsBcom}
\includegraphics[width=0.35\textwidth]{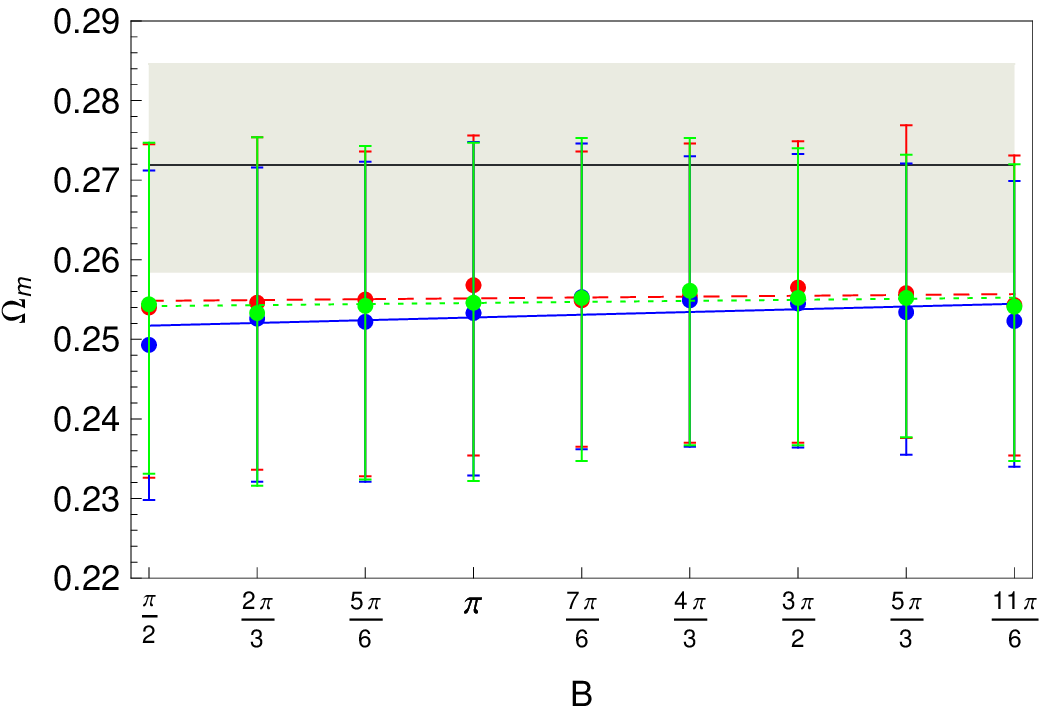}
\label{fig:omvsBcom}
\includegraphics[width=0.35\textwidth]{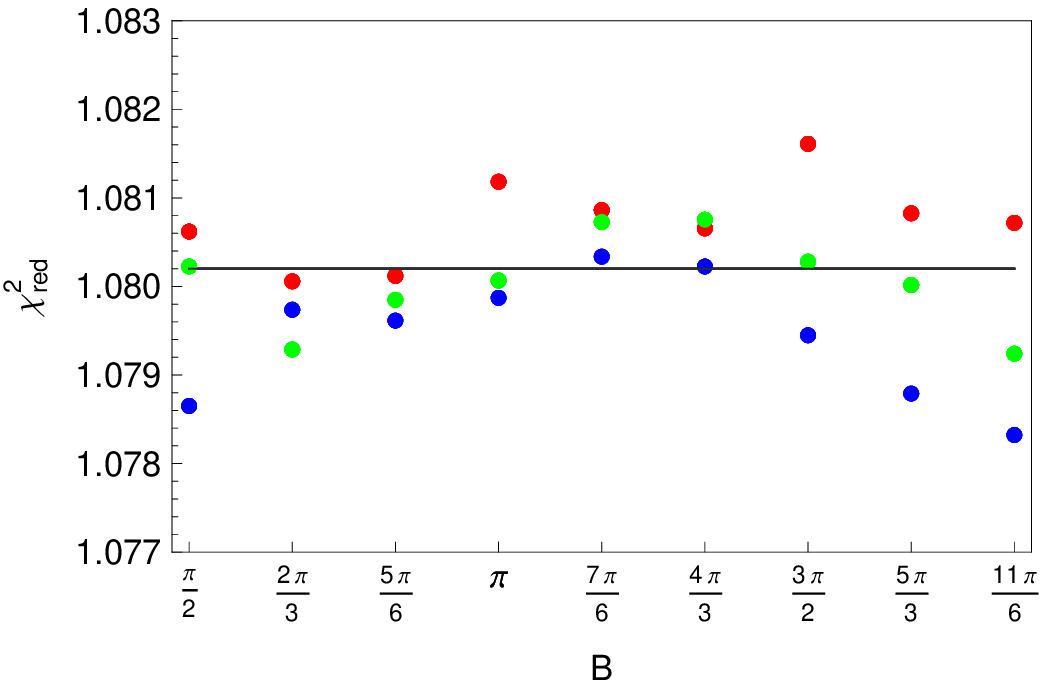}
\label{fig:chivsBcom} 
\caption{\label{fig:complots} In these
figures we plot the free parameters ${\Omega_m,A,w_0}$ and the
reduced $\chi^{2}$ value versus the fixed $B$ value for the
different models. The blue points refer to the sine model
(Eq.(6)); the red ones to the Bessel model (Eq.(7)), and the green
ones to the Struve model (Eq.(8)). The gray line in Fig.
\ref{fig:omvsBcom} and Fig. \ref{fig:chivsBcom} corresponds to
$\Lambda$CDM values. In Fig. \ref{fig:omvsBcom} the gray region
indicates the values of $\Omega_m$ within $1\sigma$ error that
describe $\Lambda$CDM.}
\end{center}\end{figure}

 Tables \ref{Sintable}, \ref{Bestable} and \ref{Strutable}
 summarize our results. As ours is a more exhaustive analysis
 than previous ones in the literature it is clear that we can
 draw stronger conclusions. The main one is that the current
data do not seem to give as strong constraints on $A$ as on the
other parameters $\Omega_m$ and $w_0$, but a clear trend in 
$A$ can be guessed, which becomes quite evident in Fig. \ref{fig:AvsBcom}. 
A fit of $A$ as a function of $B$ using a linear relation turns out to be the best one 
for the sine model; whereas, for the Bessel and Struve function a quadratic relation is preferred.

\begin{figure}\begin{center}
\includegraphics[width=0.4\textwidth]{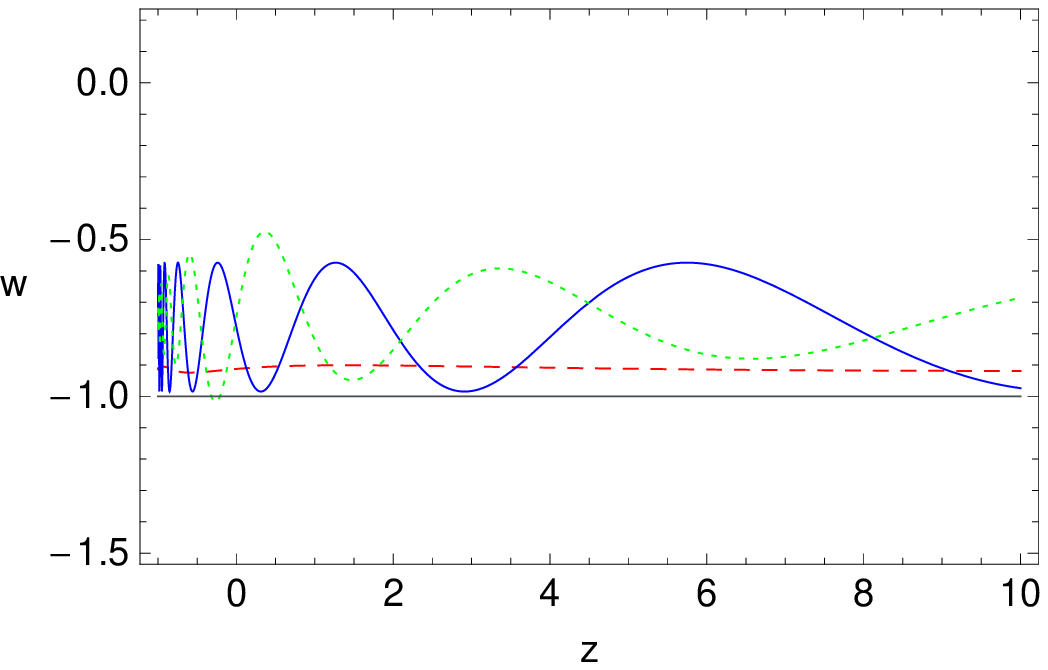}
\label{fig:wzbest}
\includegraphics[width=0.4\textwidth]{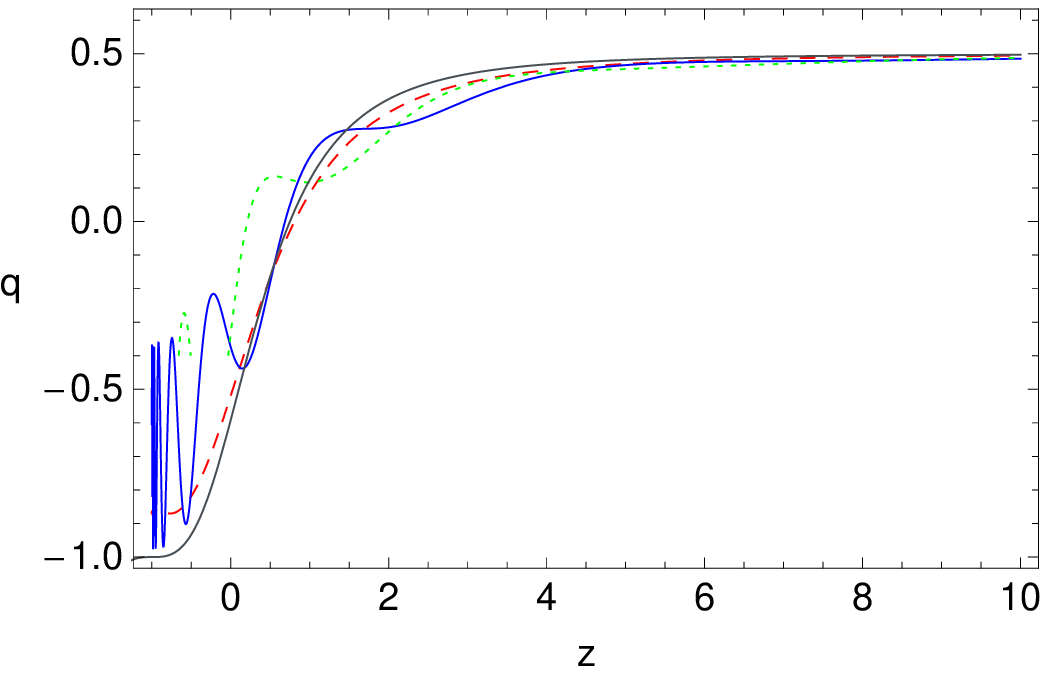}
\label{fig:qzbest} 
\caption{\label{fig:wqzbest} In these figures
we plot the redshift-variation of the EoS and the acceleration for
the best values of the parametrizations. The colors have the same
meaning of Figs. (2).}
 \end{center}\end{figure}

\begin{table}
\begin{minipage}{\textwidth}
\caption{Averaged EoS: $\overline{w}(a) =
w_{0}+\frac{A}{B}\frac{\cos(B\ln
a)-1}{\ln(a)}$.}\label{Sintable}
\hspace{-0.5cm}\resizebox*{0.49\textwidth}{!}{
\begin{tabular}{|c|c|c|c|c|c|}
\hline
\textbf{B} & $\mathbf{\boldsymbol\chi^2_{red}}$ & $\mathbf{z_{min}}$ & $\mathbf{\Omega_m}$ & \textbf{A} & $\mathbf{w_0}$ \\ \hline \hline
$\mathbf{\frac{\boldsymbol{\pi}}{2}}$ & $1.0787$ & $53.59 $ & $0.25^{+0.02}_{-0.02}$ & $0.23^{+0.37}_{-0.53}$ & $-0.95^{+0.13}_{-0.12}$  \\
$\mathbf{\frac{2\boldsymbol\pi}{3}}$ & $1.0797$ & $19.09$ & $0.25^{+0.02}_{-0.02}$ & $0.04^{+0.38}_{-0.37}$ & $-0.91^{+0.12}_{-0.13}$  \\
$\mathbf{\frac{5\boldsymbol\pi}{6}}$ & $1.0796$ & $10.02$ & $0.25^{+0.02}_{-0.02}$ & $0.10^{+0.30}_{-0.35}$ & $-0.93^{+0.13}_{-0.14}$  \\
$\boldsymbol\pi$ & $1.0799$ & $6.39$ & $0.25^{+0.02}_{-0.02}$ & $0.07^{+0.34}_{-0.38}$ & $-0.93^{+0.17}_{-0.17}$  \\
$\mathbf{\frac{7\boldsymbol\pi}{6}}$ & $1.0803$ & $4.55$ & $0.26^{+0.02}_{-0.02}$ & $-0.08^{+0.34}_{-0.41}$ & $-0.86^{+0.21}_{-0.16}$  \\
$\mathbf{\frac{4\boldsymbol\pi}{3}}$ & $1.0802$ & $3.48$ & $0.25^{+0.02}_{-0.02}$ & $-0.14^{+0.36}_{-0.35}$ & $-0.84^{+0.22}_{-0.19}$  \\
$\mathbf{\frac{3\boldsymbol\pi}{2}}$ & $1.0794$ & $2.79$ & $0.25^{+0.02}_{-0.02}$ & $-0.18^{+0.30}_{-0.34}$ & $-0.80^{+0.21}_{-0.17}$  \\
$\mathbf{\frac{5\boldsymbol\pi}{3}}$ & $1.0788$ & $2.32$ & $0.25^{+0.02}_{-0.02}$ & $-0.21^{+0.30}_{-0.27}$ & $-0.78^{+0.17}_{-0.20}$  \\
$\mathbf{\frac{11\boldsymbol\pi}{6}}$ & $1.0783$ & $1.98$ & $0.25^{+0.02}_{-0.02}$ & $-0.21^{+0.26}_{-0.22}$ & $-0.78^{+0.14}_{-0.19}$  \\ \hline
\end{tabular}}\label{tab:Sintable}
\end{minipage}
\end{table}

\begin{table}
\begin{minipage}{\textwidth}
\caption{Averaged EoS: $\overline{w}(a) =
w_{0}+\frac{A}{B}\frac{J_0(B\ln a)-1}{\ln(a)}
$.}\label{Bestable}
\hspace{-0.5cm}\resizebox*{0.49\textwidth}{!}{
\begin{tabular}{|c|c|c|c|c|c|}
\hline
\textbf{B} & $\mathbf{\boldsymbol\chi^2_{red}}$ & $\mathbf{z_{min}}$ & $\mathbf{\Omega_m}$ & \textbf{A} & $\mathbf{w_0}$ \\ \hline \hline
$\mathbf{\frac{\boldsymbol{\pi}}{2}}$ & $1.0806$ & $53.39$ & $0.25^{+0.02}_{-0.02}$ & $-0.11^{+1.00}_{-.93}$ & $-0.89^{+0.13}_{-0.14}$  \\
$\mathbf{\frac{2\boldsymbol\pi}{3}}$ & $1.0801$ & $19.09$ & $0.25^{+0.02}_{-0.02}$ & $-0.02^{+0.82}_{-0.81}$ & $-0.91^{+0.14}_{-0.13}$  \\
$\mathbf{\frac{5\boldsymbol\pi}{6}}$ & $1.0801

$ & $10.02$  & $0.26^{+0.02}_{-0.02}$ & $-0.00^{+0.60}_{-0.60}$ & $-0.90^{+0.12}_{-0.13}$  \\
$\boldsymbol\pi$ & $1.0818$ & $6.39$ & $0.26^{+0.02}_{-0.02}$ & $-0.20^{+0.63}_{-0.65}$ & $-0.86^{+0.15}_{-0.14}$  \\
$\mathbf{\frac{7\boldsymbol\pi}{6}}$ & $1.0809$ & $4.55$& $0.25^{+0.02}_{-0.02}$ & $-0.19^{+0.67}_{-0.55}$ & $-0.86^{+0.15}_{-0.17}$  \\
$\mathbf{\frac{4\boldsymbol\pi}{3}}$ & $1.0807$ & $3.48$ & $0.26^{+0.02}_{-0.02}$ & $-0.20^{+0.59}_{-0.67}$ & $-0.85^{+0.18}_{-0.17}$  \\
$\mathbf{\frac{3\boldsymbol\pi}{2}}$ & $1.0816$ & $2.79$ & $0.26^{+0.02}_{-0.02}$ & $-0.55^{+0.70}_{-0.66}$ & $-0.73^{+0.21}_{-0.22}$  \\
$\mathbf{\frac{5\boldsymbol\pi}{3}}$ & $1.0808$ & $2.32$ & $0.26^{+0.02}_{-0.02}$ & $-0.58^{+0.66}_{-0.62}$ & $-0.71^{+0.20}_{-0.23}$  \\
$\mathbf{\frac{11\boldsymbol\pi}{6}}$ & $1.0807$ & $1.98$ & $0.25^{+0.02}_{-0.02}$ & $-0.75^{+0.63}_{-0.56}$ & $-0.65^{+0.19}_{-0.23}$  \\ \hline
\end{tabular}}
\end{minipage}
\end{table}

If we pay attention to the behaviour of $w_0$, which is present value of
EoS, we see that is very well constrained. There is a slight
difference between the sine and the Bessel or Struve  cases,
but in all cases we can exclude phantom like behaviour at
$1\sigma$ level (the prior does not in principle hinder it as it rather weak). 
In Fig.\ref{fig:w0vsBcom}, we can see that the
behaviour for $w_0$ as a function of $B$ agrees with the tendency of the
amplitude $A$: as $B$ grows the value of $A$ becomes more negative so
$w_0$ moves to less negative values, see Eq.\ref{eq:modelbessel}
and Eq.\ref{eq:modelsine}. If we explore by means of fits how $w$ and $B$ are related, we find the same pattern as for $A$, 
the linear fit is preferred for the EoS with a sine form, and the 
quadratic one for the EoS with the Bessel or Struve function.

The remaining parameter, $\Omega_m$, is very well constrained and fully agrees with literature expected values, being $\Omega_{m}\simeq0.25$
and this value changes negligibly with $B$, see Fig.\ref{fig:omvsBcom}.

So, with the current data  it seems that  $B$ cannot be really constrained, that is,
all $B$ values seem to be of similar statistical validity. We have
a very slight preference for values which are different from those
chosen by other authors. 
Generally the chosen value is $B =3\pi/2$ as corresponds to the typical supernovae redshift range.
In our analysis the lowest value of  for chi-square does not correspond to that
value of $B$; for
the sine and Struve models the minimum is for $B = 11\pi/6$ which
means that we can detect oscillations within a redshift $z \sim
1.98$, comprised by our data. For the Bessel oscillating model we
have a chi-square minimum value at $B = 2\pi/3$ which will need a
redshift $z \sim 19.09$, which is outside of our observational
data range. Fig. \ref{fig:wqzbest} reflects clearly  these
behaviours: the  best values, from the statistical point of view, for the 
sine and Struve models depend allow us to detect a
complete oscillation in the range of our observational data.

\begin{table}
\begin{minipage}{\textwidth}
\caption{Averaged EoS: $\overline{w}(a) =
w_{0}+\frac{A}{B}\frac{(\pi/2)H_{-1}(B\ln
a)-1}{\ln(a)}$.}\label{Strutable}
\hspace{-0.5cm}\resizebox*{0.49\textwidth}{!}{
\begin{tabular}{|c|c|c|c|c|c|}
\hline
\textbf{B} & $\mathbf{\boldsymbol\chi^2_{red}}$ & $\mathbf{z_{min}}$ & $\mathbf{\Omega_m}$ & \textbf{A} & $\mathbf{w_0}$ \\ \hline \hline
$\mathbf{\frac{\boldsymbol{\pi}}{2}}$ & $1.0802$ & $53.39$ & $0.25^{+0.02}_{-0.02}$ & $-0.02^{+0.80}_{-0.78}$ & $-0.90^{+0.12}_{-0.14}$  \\
$\mathbf{\frac{2\boldsymbol\pi}{3}}$ & $1.0793$ & $19.09$ & $0.25^{+0.02}_{-0.02}$ & $0.17^{+0.52}_{-0.65}$ & $-0.94^{+0.14}_{-0.12}$  \\
$\mathbf{\frac{5\boldsymbol\pi}{6}}$ & $1.0798$ & $10.02$  & $0.25^{+0.02}_{-0.02}$ & $0.05^{+0.51}_{-0.55}$ & $-0.91^{+0.14}_{-0.15}$  \\
$\boldsymbol\pi$ & $1.0801$ & $6.39$ & $0.25^{+0.02}_{-0.02}$ & $0.01^{+0.44}_{-0.50}$ & $-0.91^{+0.15}_{-0.15}$  \\
$\mathbf{\frac{7\boldsymbol\pi}{6}}$ & $1.0807$ & $4.55$& $0.26^{+0.02}_{-0.02}$ & $-0.14^{+0.51}_{-0.44}$ & $-0.86^{+0.16}_{-0.18}$  \\
$\mathbf{\frac{4\boldsymbol\pi}{3}}$ & $1.0808$ & $3.48$ & $0.26^{+0.02}_{-0.02}$ & $-0.20^{+0.56}_{-0.45}$ & $-0.84^{+0.18}_{-0.20}$  \\
$\mathbf{\frac{3\boldsymbol\pi}{2}}$ & $1.0803$ & $2.79$ & $0.26^{+0.02}_{-0.02}$ & $-0.21^{+0.49}_{-0.49}$ & $-0.83^{+0.21}_{-0.21}$  \\
$\mathbf{\frac{5\boldsymbol\pi}{3}}$ & $1.0800$ & $2.32$ & $0.26^{+0.02}_{-0.02}$ & $-0.38^{+0.45}_{-0.50}$ & $-0.74^{+0.22}_{-0.19}$  \\
$\mathbf{\frac{11\boldsymbol\pi}{6}}$ & $1.0792$ & $1.98$ & $0.25^{+0.02}_{-0.02}$ & $-0.36^{+0.39}_{-0.36}$ & $-0.75^{+0.18}_{-0.19}$  \\ \hline
\end{tabular}}
\end{minipage}
\end{table}

Nevertheless, focusing our attention on Fig. \ref{fig:chivsBcom},
we can see that for almost all the values of the frequency, $B$,
the values of chi-square of the sine parametrization are
smaller than those of others parametrizations. If we take this
into account and the fact that the best value of this
parametrization corresponds to a $B=11\pi/6$, we could say that
observational data shows a preference for a periodic EoS with a
small period.

At the same time, if we look at the variation of the
acceleration parameter (Fig.\ref{fig:qzbest}) in the recent past we
can observe that it has recently peaked and is slowing down at present as
have been pointed in \cite{Shafieloo2009}.

Moving to the statistical side of the analysis, we have to argue
if the proposed models are reasonably good or not, and above all if they can
compete with the the concordance $\Lambda$CDM model.
From our results we conclude an oscillating pattern in the dark energy is an admissible \textit{possibility}, as there
is so far no concluding evidence against it; and from some statistical perspectives they even represent a better option
that its main competitor,  $\Lambda$CDM. Tables \ref{stat} 
 summarizes our findings on the statistical side.

\begin{table}
\begin{minipage}{\textwidth}
\caption{Statistical criteria}\label{stat}
\hspace{-0.5cm}\resizebox*{0.49\textwidth}{!}{
\begin{tabular}{|c|c|c|c|c|c|c|}
\hline
Model & $\mathbf{\boldsymbol\chi^2}$ & $\mathbf{\boldsymbol\chi^2_{red}}$ & $\Delta$AICc & $\Delta$BIC & $\Delta$RICc & $\Delta$DICc \\ \hline \hline
$\Lambda$CDM & $507.7003$ & $1.0802$ & $0$ & $0$ & $0$ & $0$ \\
Sinus & $504.6550$ & $1.0783$ & $0.9975$ & $9.2687$ & $-1.0282$ & $1.9145$ \\
Bessel & $505.4669$ & $1.0801$  & $1.8094$ & $10.0806$ & $-0.2163$ & $2.0344$ \\
Struve & $505.0848$ & $1.0792$ & $1.4273$ & $9.6985$ & $-0.5984$ &
$2.0023$\\ \hline
\end{tabular}}
\end{minipage}
\end{table}

If we take a look at the reduced chi-square values, we can see
that all the models have lower values than the $\Lambda$CDM one,
even if the differences are really small. In particular, the sine
model seems the best one. If we take a look to the
$\mathrm{AIC_{c}}$ values, we see that the sine model is just in
a border line position if we consider the limit we discussed in
Section \ref{sec:testmodel}, i.e. $\Delta \mathrm{AIC} \gtrsim 1$.
On the contrary, we should absolutely reject  the Bessel and Struve models. 
But the situation changes when considering
$\mathrm{RIC_{c}}$: we know that it has a small penalty than
$\mathrm{AIC_{c}}$ when $N>>k$ as it is in our case, where we have
$N=472$ and $k=3$. And we see that its values favour the
oscillating models with respect of $\Lambda$CDM; they are even
negative, meaning that the $\mathrm{RIC_{c}}$ of oscillating
parametrizations are better.

The situation reverses again when moving to $\mathrm{BIC}$; if we
compare the values obtained with the Jeffreys' scale, we should
conclude that there is an \textit{almost decisive} evidence
against oscillating patterns. But we have to remember that
$\mathrm{BIC}$ has some problem when facing degeneracies
between parameters, like the behaviours of the amplitude and of
$w_{0}$ seem to reflect.

These degeneracies would well be the reason of the ``bad results" offered by $\mathrm{BIC}$,
which is also challanged by another criterion, $\mathrm{DIC}$, which looks more
favourable with oscillating dark energy. Since 
$\mathrm{DIC}$ relies also on the  Bayesian approach, we can apply  Jeffreys'
scale to it in the same fashion as above, and from this we conclude there is not a
significant evidence against oscillations.

\section{Conclusions}

In this paper we have performed a quite detailed analysis towards the detection of
oscillating patterns in the dark energy EoS. We have considered
different phenomenological models, starting
from the original sine models and then introducing two new ones, based on the 
special functions. Those new  models differ from the sine one mainly 
because they show a damped amplitude in the oscillations when moving into the past.

Compared to prior works devoted to oscillating dark energy we
can highlight that fact that instead of fixing the frequency parameter to a particular value, 
we have explored a discrete set  of frequency values. 

We have also introduced novelty with respect of the datasets used, as we take direct
measurements of the Hubble factor and GRB luminosity data, which are arguably 
two directions of improvement as the sensitivity to oscillating patterns gets increased
and a wider redshift range comes into play. Our theoretical setup has
been complemented with  a detailed statistical analysis using different model selection tools.

Numerical results show that while parameters like the matter
content, $\Omega_{m}$ and the present value of dark energy EoS,
$w_{0}$ can be constrained very well, this is not true for the
amplitude and the frequency. In particular, between the amplitude 
$A$ and $w_{0}$ seems to be working a degeneracy that cannot be
solved. About the frequency, we can say that chi-square values are
not really in favour of a particular value, being all in a very
narrow range. But the best values favour values of the frequency
which mean oscillation detectable inside the present redshift
range of SNeIa.

The statistical analysis seems does not seem to provide such a conclusive answer 
as desirable, though we think that oscillations can be considered as a
possible alternative to a $\Lambda$CDM model for addressing the
well know problems it suffers from (as review in the
Introduction).  All but one ($\mathrm{BIC}$) of the statistical criteria used  considered here are keen to the 
possible detection of oscillations in the EoS, there is even one of them   ($\mathrm{RIC}$) which seems to favor an oscillating pattern
against a cosmological constant.

\section*{Acknowledgements}
We are grateful to D. Yonetoku for providing us the GRB data and to E. Komatsu and M. Hicken for useful comments.
We also wish to thank IZO-SGI SGIker (UPV/EHU, MI-
CINN, GV/EJ, ESF), and in particular E. Ogando and T. Mercero, for
technical and human support. R.L., V.S. and I.S. are  sustained by the Basque Government through grant GIU06/37. 
R.L. and I.S. have additional support from
the Spanish Ministry of Science and Innovation through grant FIS2007- 61800.

\end{document}